\RequirePackage{fixltx2e}
\documentclass[aps,preprint,showpacs,superscriptaddress,12pt]{revtex4-1} 

\usepackage[normalem]{ulem}
\usepackage{xcolor}

\usepackage{bm}
\usepackage{graphicx}
\usepackage{latexsym}
\usepackage{amsmath}
\usepackage{hyperref}
\usepackage{pxfonts}
\usepackage{cancel}
\usepackage{bbding}
\usepackage{subfig}
\usepackage{float}
\usepackage{subfloat}
\usepackage{caption}
\usepackage[section]{placeins}
\usepackage{csquotes}
\DeclareGraphicsExtensions{.pdf,.png,.jpg}

\begin{document}

\title{\Large{Non-linear Ion-Wake Excitation by Ultra-relativistic Electron Wakefields}}

\author{Aakash A. Sahai} 
\email{aakash.sahai@gmail.com}
\affiliation{Department of Electrical Engineering, Duke University, Durham, NC, 27708 USA}
\author{Thomas C. Katsouleas}
\affiliation{Department of Electrical Engineering, Duke University, Durham, NC, 27708 USA}

\begin{abstract}
The excitation of a non-linear ion-wake by a train of ultra-relativistic electron bubble wake  \cite{Akhiezer-Polovin}-\cite{cavitation-beam-expt} is modeled. The ion-wake is shown to be a driven non-linear ion-acoustic wave in the form of a cylindrical ion-soliton \cite{Maxon-cyl-soliton}-\cite{KdV-non-linear-ion-waves}. The phases of the oscillating radial electric fields of the slowly-propagating \cite{Vlasov} electron bubble is asymmetric in time and excites time-averaged inertial ion motion radially. The electron compression in the back of the bubble sucks-in the ions and the space-charge within the bubble cavity expels them, driving a cylindrical ion-soliton structure with on-axis and bubble-edge density-spikes \cite{Pukhov-laser-bubble}\cite{Lu-bubble-regime}. Once formed, the channel-edge density-spike is driven radially outwards by the thermal pressure of the wake energy \cite{cyl-soliton-observation}. Its channel-like structure due to the flat-residue left behind by the propagating ion-soliton, is independent of the energy-source driving the bubble \cite{cavitation-laser}\cite{cavitation-beam} electron wake. We explore the use of the partially-filled channel formed by the cylindrical ion-soliton for a novel regime of positron acceleration \cite{positron-accln-2001}. OSIRIS PIC \cite{osiris-code-2002} simulations are used to study the ion-wake soliton structure, its driven propagation and its use for positron acceleration.

\end{abstract}

\maketitle 

\section{Introduction}
Plasma ions are generally assumed to be stationary in the theory of ultra-relativistic non-linear plasma electron waves excited as wakefields of a laser or particle beam \cite{Akhiezer-Polovin}-\cite{Lu-bubble-regime}. Important exceptions to this occur when the drive beam (or a witness beam being accelerated) in a plasma wakefield accelerator becomes denser than the background plasma density by a factor larger than the ion to electron mass ratio \cite{ion-motion-intense-beam}\cite{ion-motion-beam-emittance}. Such is the case for the parameters of ultra-low emittance future collider designs at the TeV energy scale. Ion motion also becomes important in the wake train behind the initial wake oscillation or bubble. Understanding the long-term ion behavior \cite{cavitation-laser-expt}\cite{cavitation-beam-expt} is important to determine the state of the plasma for succeeding pulses in a future plasma collider of high repetition rate \cite{long-term-wake}\cite{hot-plasma-wake}. We also show that the resulting ion-wake channel formed is of interest for positron wakefield acceleration.

\begin{figure}[ht!]
	\begin{center}
   	\includegraphics[width=\columnwidth]{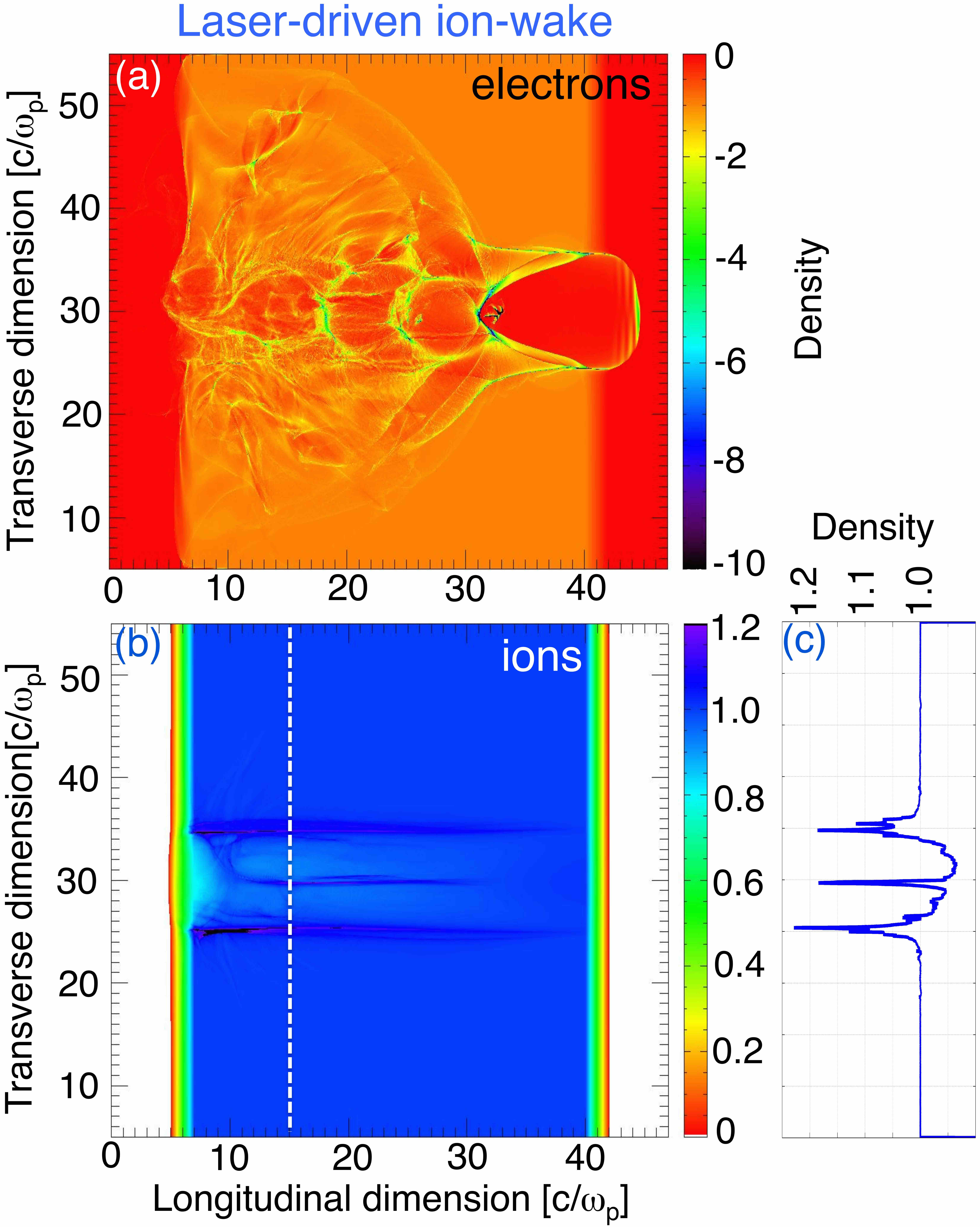}
	\end{center}
\caption{ \scriptsize {\bf Laser driven non-linear ion-wake at early time (t = 46$\omega_{pe}^{-1}$ = 0.17 $f_{pi}^{-1}$) in $m_{i}=m_p=1836 ~ m_e$ plasma}. (a) Electron bubble wakefields in cartesian coordinates (fixed-box) with $\frac{\omega_0}{\omega_{pe}}=10$ driven by a matched laser pulse (vector potential $a_0=4$ and frequency $\omega_0$) with $R_B \simeq 4\frac{c}{\omega{pe}}$. (b) Non-linear ion-wake in the form of a cylindrical ion-soliton of radius $\simeq 4\frac{c}{\omega{pe}}$ excited behind the bubble electron wake in a proton plasma. (c) Transverse ion-density profile at $z=15~c/\omega_{pe}$.  Notice that the ion density perturbation in this excitation phase is still building up and is a fraction of the background ion density, $\frac{\delta n_i}{n_0} < 1$.}
\label{fig1:ion-wake-laser}
\end{figure}

In this paper, we model the excitation of a nonlinear ion-wake in the trail of an electron plasma wake in the bubble regime, driven by either relativistically intense laser or particle beam energy sources. The ion-wake is shown to be a non-linear ion-acoustic wave in the form of a cylindrical ion-acoustic soliton. Its characteristics are similar to the solutions of the cylindrical Korteweg - de Vries equation (cKdV) \cite{Maxon-cyl-soliton}-\cite{KdV-non-linear-ion-waves}. However, the ion-wake shown here is different from a cylindrical-KdV soliton due to several reasons: The bubble wake electron oscillations do not thermalize into an isothermal plasma; the ion-wake soliton breaks up into N-solitons as it evolves and there is an ion-density spike on the axis. The soliton propagates radially outwards leaving behind a flat residue resulting in a partially filled ion-wake channel. The dynamics of the ion wake is shown to have two distinct phases. The excitation phase where the time asymmetry of the radial electric fields of the bubble excite the ions into a soliton-like structure. At later times, the propagation phase where the soliton is driven outwards by the pressure of thermalized electrons after the electron oscillations undergo phase mixing \cite{phase-mixing-longitudinal}. 

\begin{figure}[h!]
	\begin{center}
   	\includegraphics[width=\columnwidth]{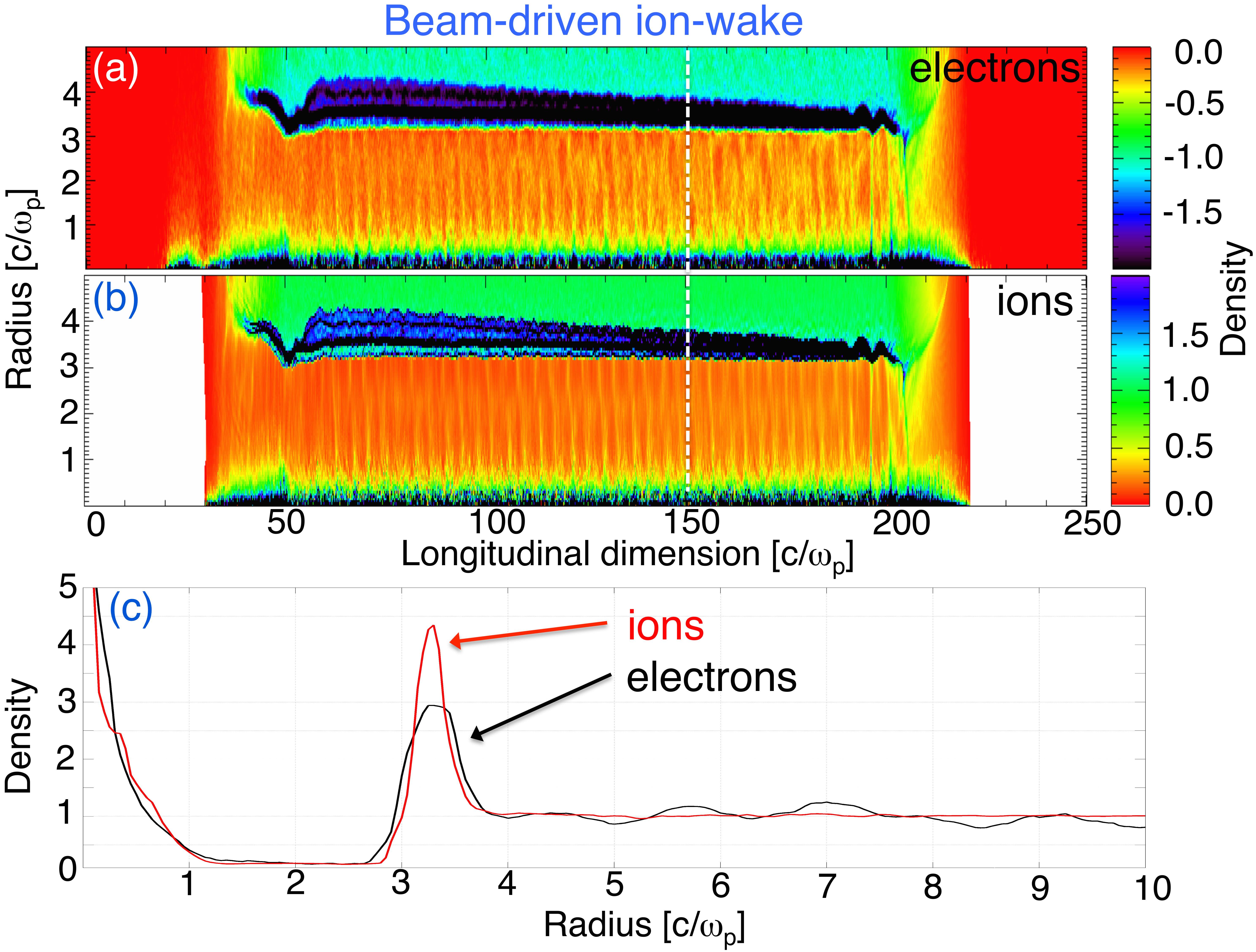}
	\end{center}
\caption{ \scriptsize {\bf Electron beam-driven non-linear ion-wake at late time (t = 460 $\omega_{pe}^{-1} = 1.7 f_{pi}^{-1}$) in $m_{i}=m_p=1836 ~ m_e$ plasma}. (a) Beam-driven ion-wake electron density in cylindrical coordinates (fixed-box). The beam parameters are $n_b=5n_0$, $\sigma_r=0.5c/\omega_{pe}$, $\sigma_z = 1.5 c/\omega_{pe}$, $\gamma_b=38,000$, these beam-plasma parameters are quite similar to \cite{cavitation-beam}. (b) Corresponding ion density in cylindrical coordinates (fixed-box). Note the N-soliton formation in the ion-density. This is seen as the ion-wake evolves further, for instance at  $z=60~c/\omega_{pe}^{-1}$. (c) Radial electron and ion density profile at $z=150~c/\omega_{pe}$. }
\label{fig2:ion-wake-beam}
\end{figure}

Representative PIC simulation results in Fig.\ref{fig1:ion-wake-laser} and Fig.\ref{fig2:ion-wake-beam} illustrate the salient features of the non-linear ion-wake. Figure\ref{fig1:ion-wake-laser} shows the excitation phase at an early time when the bubble wake train is executing ordered oscillations and its fields have begun to excite the ion-wake structure seen in Fig.\ref{fig1:ion-wake-laser}(b),(c). The radial oscillations later phase mix and the electrons thermalize, their thermal pressure then drives the ion-soliton outwards as shown in  Fig.\ref{fig2:ion-wake-beam}. As we show below, it is the longitudinal or time asymmetry of the radial electron wakefields that excites the ion soliton which propagates leaving behind a partially-filled channel shown in the PIC simulations in Fig.\ref{fig1:ion-wake-laser} and Fig.\ref{fig2:ion-wake-beam}.

In section \ref{ion-wake-model-considerations} we present considerations and assumptions made to derive the ion-wake model. In section \ref{radial-ion-wave} using the fluid equations for ion dynamics we model the non-linear ion-acoustic waves as a driven cylindrical ion soliton. The driven wave models are shown to have two distinct phases: the excitation phase and the propagation phase. We present a model for the excitation of the ion-soliton by the electron bubble radial fields in section \ref{ion-soliton-excitation-phase} and verify it using simulation results. In section \ref{ion-soliton-propagation-phase} we present simulation results showing the dynamics of propagation of the ion-soliton driven by a radial temperature gradient. Finally, in section \ref{ion-soliton-positron-wake} we analyze the properties of a positron beam driven wakefield in an ion-wake channel. 

\section{Considerations for modeling the Non-linear ion-wake}
\label{ion-wake-model-considerations}

We recognize that to study the time evolution of a wake-excited plasma in order to establish the duration over which it relaxes to  thermal equilibrium both collisional and collision-less dynamics have to be considered along with the physics of recombination modes such as electron-ion recombination. However, in this work the dynamics of plasma is modeled under the collision-less approximation. Thus, diffusion is not important during the timescales over which the ion-wake is studied. We do not discuss recombination except mentioning that the ``afterglow" is dominated by volume recombination while localized effects cannot be ruled out.

In order to formally establish the difference between the density wave processes that occur over collision-less timescale in contrast to the ones that start dominating under collisions, we show the assumptions made to arrive at the dynamics of diffusion. The process of diffusion is modeled with a parabolic partial differential equation which is deduced from the ion-fluid equations under the assumption that the inertial response of the ions is much faster than the collisional timescales. 

The effect of collisions is introduced as a drag force, $m_in ~ \nu_{coll} ~ \langle \vec{v}_i \rangle$. The collisional drag force modifies the ion-fluid equation of motion as, $m_i n_i\frac{\operatorname d\vec{v}_i}{\operatorname dt} = mn\left(\frac{\partial \langle \vec{v}_i \rangle}{\partial t} + \langle \vec{v}_i \rangle \vec{\nabla} \cdot \langle \vec{v}_i \rangle \right) = \pm e ~ n\vec{E} -  \vec{\nabla}\mathcal{P}_e - m_i n_i ~ \nu_{coll} ~ \langle \vec{v}_i \rangle $ where $\nu_{coll}$ is the average electron-ion collision frequency and is obtained from the mean free path. Diffusion of plasma is thus driven by the charge-separation field, $\vec{E}$ and the thermal pressure, $\mathcal{P}_e$ while being impeded by the collisional drag. Upon ignoring the inertia of the ions, $\frac{\partial \vec{v}_i}{\partial t} = 0$, the equation for the ion velocity by diffusion in an isothermal plasma is, $\langle \vec{v}_i \rangle =  \pm \frac{e}{m ~ \nu_{coll}} ~ \vec{E} - \frac{k_BT_e}{m ~ \nu_{coll}} \frac{\vec{\nabla}n_i}{n_0}$. The characteristic parameter of diffusion is the diffusion coefficient or diffusivity $D = \frac{k_BT_e}{m ~ \nu_{coll}}$ and mobility $\mu = \frac{e}{m ~ \nu_{coll}}$ which depend upon the collision frequency. Using the gradient of the velocity in the continuity equation and ignoring the mobility $\mu$, leads to a Fick's law diffusion equation, $\nabla^2\frac{n_i}{n_0} \propto \frac{\partial }{\partial t} \frac{n_i}{n_0}$; characteristic of a parabolic equation. 

When the mobility is retained, the fluid equation is a moment of the Fokker-Planck equation which is the kinetic model of the collision-driven drift and diffusion. The diffusion equation thus cannot support wave-like solution because such solutions are characteristic of a hyperbolic partial differential equation. 

The solutions of linear and non-linear diffusion equations show the evolution of density profile by diffusion and can be obtained using the self-similar formulation. The self-similar solutions show the spatial and temporal evolution of the density to be exponentially decaying. In the non-linear case, the density can have a sharp-front as it decays. However, a soliton-like propagating solution cannot be described with diffusion equation. Hence, the cylindrical ion-soliton presented here is not diffusion but a wave phenomenon.

The electron bubble wake is excited by a sub-wavelength impulse of an ultra-short driver. In contrast, the ion-wake is excited as the ions undergo sustained interactions with the bubble fields within the spatial extent of the wake over several plasma electron oscillations. This happens because the electron wake-plasmon oscillations \cite{wakefield} have a near speed-of-light phase-velocity ($\beta_{\phi} \simeq 1$) but negligible group-velocity \cite{Vlasov} $\beta_g\approx 3 v_{th}^2/c^2$ (in the 1-D limit), where, $v_{th} \simeq \sqrt{k_BT_e/m_e}$ is the mean electron thermal velocity of the background plasma. Therefore a slowly-propagating train of coupled electron plasmons is excited in a cold collision-less plasma \cite{Vlasov}. A large difference between phase-velocity and the group velocity of the electron oscillations allows sustained field-ion interactions. It should be noted that high phase-velocity plasma electron waves are possible only in a cold plasma with appropriate density, $n_0$ that allows near speed-of-light propagation of the energy sources, $\beta_{es} \simeq 1 \approxeq \beta_{\phi}$. Ion-soliton modeled here is assuming a significant difference between the phase-velocity and the group velocity of the plasma-electron waves.

A time symmetric electron wakefield would excite time symmetric ion oscillations where the ion velocities average to zero. However, the bubble wake is asymmetric in time as the back of the bubble electron compression is a small fraction of the length of electron cavitation. The electron oscillations become non-linear at high driver intensities as all the interacting electrons are displaced radially, $\delta n_e/n_0 > 1$, forming a non-linear bubble-shaped electron spatial structure enclosing ions in its cavity. The wakefields excited in the bubble are useful for accelerating electrons \cite{cavitation-beam}\cite{cavitation-laser-expt}\cite{cavitation-beam-expt}. High intensities also lead to fields that can directly drive the plasma electrons to velocities near the speed-of-light. This occurs when for a laser pulse $a_0 \geq 1$ and an electron beam $\frac{n_b}{n_0}\left(\frac{r_b}{c/\omega_{pe}}\right)^2\geq 1$ where $a_0$ is the peak normalized laser vector potential, $n_b, r_b$ the peak beam density and radius. The radially expelled electrons oscillate radially under the force of the plasma ions. These oscillations are excited over plasma electron oscillation timescales, $2\pi\omega_{pe}^{-1}$ ($\omega_{pe}=\sqrt{4\pi n_0 e^2/\gamma_em_e}$) where $\gamma_e\beta_em_ec$ is the temporally anharmonic relativistic electron quiver momentum. The normalized quiver momentum of the electrons in the bubble-oscillations is relativistic $\gamma_{\perp}\beta_{\perp} \geq 1$ and the quiver frequency is $\omega_{\perp}=\omega_{pe} \left( \frac{\beta_{\phi}^2}{\gamma(1-\beta_{\phi}^2)} \right)^{1/2}$ \cite{Akhiezer-Polovin}. 

We show that non-linear ultra-relativistic electron wakefields interacting with the plasma ions lead to the excitation of a non-linear ion-wake. The non-linear ion-wake $\delta n_i/n_0 > 1$ in Fig.\ref{fig1:ion-wake-laser} and Fig.\ref{fig2:ion-wake-beam} is excited over timescales $\gg 2\pi\omega_{pe}^{-1}$ in the trail of a bubble-wake train. By shaping the energy source it can be matched or guided to excite a long train of nearly identical plasmons, Fig.\ref{fig3:bubble-train}. Since it is the electric field ${\bf E}_{wk}$ of a nearly stationary bubble plasmon that excite collective ion-motion we model the ion dynamics in a single bubble. Using the single bubble ion dynamics, Fig.\ref{fig5:bubble-ion-dynamics} we model the ion-wake over the whole bubble-train spanning several hundred plasma skin-depths ($c/\omega_{pe}$).

The wake-plasmon energy density ($\mathcal{E}_{wk} = 0.5 (e\lvert {\bf E}_{p}\rvert/(m_ec\omega_{pe}))^2 ~ m_e c^2 ~ n_0$, where ${\mathbf E}_{p}$ is the wakefield amplitude) is continually partitioned between the field energy and the coherent electron quiver kinetic energy. In our model we do not include heavy beam-loading of the bubble electron wake. Under heavy beam-loading the bubble field energy is efficiently coupled to the kinetic energy of the accelerated beam. In this scenario the bubble collapses and the magnitude of the ion-wake is smaller. The decoherence of the ordered electron quiver to random thermal energy, $\mathcal{E}_{wk} \rightarrow k_BT_{wk}$ due to the phase-mixing \cite{phase-mixing-longitudinal} of individual electron trajectories caused by the non-linearities and inhomogeneities is further stimulated by the ion motion. The details of the thermalization of the wake electrons under ion motion is beyond the scope of this paper. It is over these timescales upon thermalization that the steepened ion-density expands outwards radially as a non-linear ion-acoustic wave driven by the electron thermal pressure. The energy transfer process observed here is a coupling from the non-linear plasma electron-mode to a non-linear ion-acoustic mode \cite{lte-napac-2013}. We also observe energy coupling to the bow-shock which is formed behind the bubble, Fig.\ref{fig5:bubble-ion-dynamics}.

The ion-wake is a partially-filled channel with sub-skin-depth density-spikes on-axis and at the bubble-edge located at the bubble-radius, $R_{B}$ \cite{Pukhov-laser-bubble}\cite{Lu-bubble-regime} of several $c/\omega_{pe}$. The ion accumulation in both the density-spikes is many times the background density as shown in Fig.\ref{fig1:ion-wake-laser}(c) and Fig.\ref{fig2:ion-wake-beam}(c). The wake energy is coupled to the ions and to the thermal energy \cite{long-term-wake} of the de-cohering wake electrons. The interplay between the electron thermal energy and ion kinetic energy then form a cylindrically symmetric non-linear ion-acoustic wave which further empties the near-void region of the ion-wake channel. 

The time-scale of dissipation of the ion-wake and relaxation of the plasma distribution to $v_{th}/c\sim 0$ sets an upper limit on the repetition-rate \cite{hot-plasma-wake} of the future plasma colliders. It is well known that the ion acoustic wave is damped by collisions and ion-wave Landau damping. The ion dynamics also opens important questions on the design the plasma container walls in a future high-repetition rate accelerator because if a significant flux of ions strikes the wall it can cause significant damage. It was suggested that the wakefield energy in the plasma wave could be replenished and sustained \cite{beam-wakefield-SLAC} by a train of energy source in order to achieve high repetition rate. However, as shown in this paper due to ion motion this is not possible.

We explore the use of the ion-wake channel for positron-beam driven positron acceleration in a novel and relevant \enquote{suck-in} regime \cite{positron-accln-2001} where the positron beam radius $r_{pb} \gtrsim c/\omega_{pe}$. Such channels are also promising \cite{hollow-channel-chiou}\cite{hollow-acclerator-1998}\cite{hollow-accelerator-2013} for exciting electron-wakefields with independent transverse and longitudinal field spatial structures. 

The formation of plasma channels which are vaguely similar to the cylindrical soliton shown here but are excited by significantly different processes have been shown previously. These processes are: using a collimated laser with annular profile \cite{milchberg-axicon-channel}\cite{bessel-beam-2011}, by an electrical discharge \cite{hollow-channel-discharge}, by linear wakes in the self-modulated regime \cite{ting-smlwfa-channel}\cite{ponderomotive-pdpwfa-wake-channel} and by the ponderomotive force of a short-pulse laser-driven linear wake \cite{ponderomotive-linear-wake-channel}. In the linear electron wake regime the wakefields are symmetric so they average out over the electron oscillation period and only the second-order ponderomotive force ($\bf{F^{wk}_p} \approx -{\bf \nabla} \langle{\bf E_{wk}}\rangle^2$) drives the ion-motion ($m_{i}\frac{\partial^2 r_{i}}{\partial \tau^2} \simeq F^{wk}_p$) \cite{ponderomotive-pdpwfa-wake-channel}. Spatially, in the linear-wake regimes the radial ion excitation is limited to around $c/\omega_{pe}$ and the ion-density perturbation is linear in the background plasma density, $n_0$ \cite{ponderomotive-linear-wake-channel}\cite{ponderomotive-pdpwfa-wake-channel}. Additionally, in the earlier studies no ion acoustic-waves have been examined. 

\section{Ion-wake as a driven \\ radial Ion-acoustic wave}
\label{radial-ion-wave}

To develop insight into the ion wake physics, we consider the 1-D simplified dispersion relation of the ion-acoustic plane waves, 

\begin{align}
\omega^2 = \frac{c_s^2k^2}{ 1 ~ + ~ (c_s/\omega_{pi})^2 ~ k^2 }
\label{ion-wave-dispersion-relation}
\end{align}

\noindent where, $\omega_{pi}=\omega_{pe}\sqrt{m_e/m_{i}}$ and $c_s = \sqrt{\Upsilon k_BT_{wk}/m_{i}}$ under the collision-less condition, $T^i_{wk}\ll T^e_{wk}$ and $\Upsilon = 1 + 2/f$ is the adiabatic index with $f$ being the degrees of freedom of the ions. During the initial phase of the interaction between the bubble fields and the ions, the ion inertia dominates their motion. At times when the fields of the bubble interact with the ions, the inertia of ions leads to very small spatial displacement scales and so $k \rightarrow \infty$ so the term $k (c_s/\omega_{pi}) \gg 1$ where $c_s/\omega_{pi} = \lambda_{De} =\sqrt{\frac{k_BT_e}{4\pi e^2 n_0}}$ is the Debye wavelength. Thus in this phase $\omega = \omega_{pi}$ and the ion-soliton spike at $R_B$ grows over the plasma frequency timescale, $2\pi \omega_{pi}^{-1}$. At initial times the plasma is also cold so $c_s \simeq 0$. 

When the ions gain significant momentum and start oscillating over larger spatial scales in response to the electron dynamics then $k\lambda_{De} \ll 1$. In this phase the acoustic wave propagation becomes dispersion-less with $\omega = k c_s$. However, when the ion density perturbation undergoes steepening in a large amplitude acoustic wave forming a spike over much smaller spatial scales, $k$ becomes large and dispersion becomes important. The ion-acoustic dispersion relation as $k$ increases has higher-order corrections in $k$, $\omega = c_s k ~ - ~ \frac{c_s}{2} ~ \left( \frac{c_s}{\omega_{pi} } \right)^2 ~ k^3$. Also, as the ions are heated at later time $T_i$ increases and modifies the sound speed to $c_s = \sqrt{ k_B (\Upsilon_e T_e + \Upsilon_i T_i) / m_{i}}$. Thus when the ions are heated the sound speed lies between $\sqrt{ k_B (\Upsilon_e T_e + \Upsilon_i T_i) / m_{i}}$ and $\sqrt{\Upsilon_e k_B T_e / m_{i}}$.

\subsection{Linearized ion-acoustic wake}

During the time when the oscillations in the electron plasma-wave wake are undergoing phase-mixing the wakefields still persist and the ion motion is driven by the fields as well as the thermal pressure. We show under the linear approximation how an ion-wake differs from a freely propagating ion-acoustic wave. It should be noted that as the wake electrons phase-mix, the plasma is not isothermal in this case. The linear ion-acoustic wave can be obtained by perturbative expansion of ion density, $n_i$ and ion fluid velocity $v_i$ in the zeroth-order ion fluid continuity equation, $n_0\mathbf{\nabla}\cdot\mathbf{v}^{(1)}_i + \frac{\partial n^{(1)}_i}{\partial t} = 0$. Taking a partial derivative with time, $\mathbf{\nabla}\cdot\frac{\partial \mathbf{v}^{(1)}_i}{\partial t}  + \frac{\partial^2 }{\partial t^2} \frac{n^{(1)}_i}{n_0} = 0$. The ion-fluid equation of motion where the electron temperature has a spatial gradient is, $m_i\frac{\partial \mathbf{v}^{(1)}_i}{\partial t} = eZ_i\mathbf{E}_{wk} - \Upsilon k_BT_e\mathbf{\nabla}\frac{n^{(1)}_i}{n_0} - \Upsilon k_B\frac{n^{(1)}_i}{n_0}\mathbf{\nabla}T_e$. Upon substituting the equation of motion in the time-derivative of the linearized continuity equation, $\mathbf{\nabla}\cdot\left(\frac{eZ_i}{m_i}\mathbf{E}_{wk} - \frac{\Upsilon k_BT_e}{m_i}\mathbf{\nabla}\frac{n^{(1)}_i}{n_0} - \frac{ \Upsilon k_B}{m_i} \frac{n^{(1)}_i}{n_0}\mathbf{\nabla}T_e \right)  + \frac{\partial^2 }{\partial t^2} \frac{n^{(1)}_i}{n_0} = 0$. Thus, a driven ion-acoustic wave linearized to the first-order in density perturbation has the form, 

\begin{align}
\nonumber \left(\frac{\partial^2 }{\partial t^2} - c_s^2\nabla^2\right) & \frac{n^{(1)}_i({\mathbf r},t)}{n_0} \\
& = - \frac{eZ_i}{m_i}\mathbf{\nabla}\cdot\mathbf{E}_{wk}({\mathbf r},t) \biggr\rvert_{\mathrm{wake}} + \frac{ \Upsilon k_B}{m_i}\frac{n^{(1)}_i}{n_0}\nabla^2T_e \big\rvert_{\mathrm{thermal}}
\label{ion-wave-equation-first-order}
\end{align}

\begin{figure}
	\begin{center}
   	\includegraphics[width=\columnwidth]{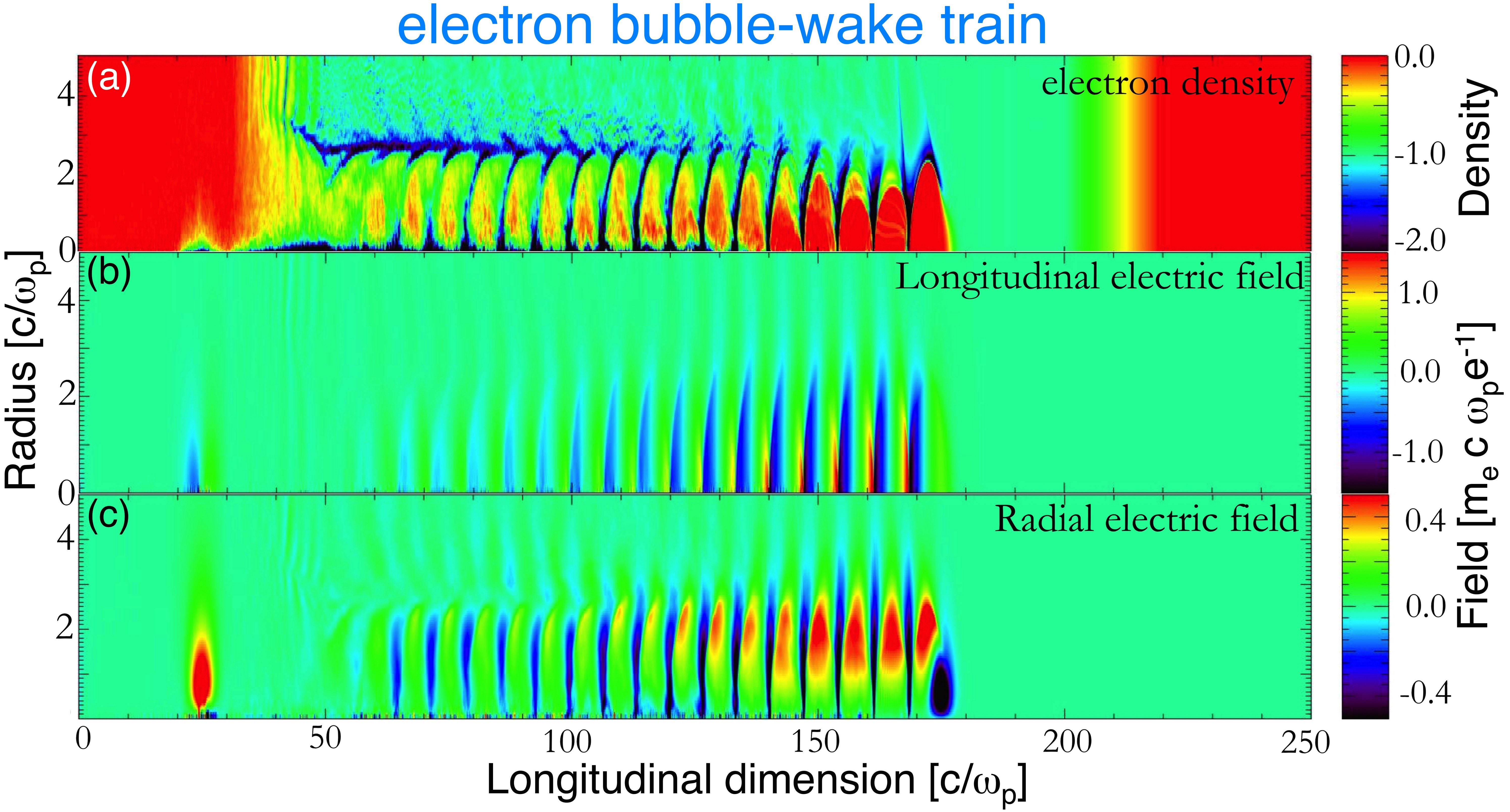}
	\end{center}
\caption{ \scriptsize { \bf Bubble-wake train behind an ultra-relativistic electron beam under the condition $\beta_g(bubble) \ll \beta_{\phi}(bubble) \simeq \beta_{beam}$ with beam-plasma parameter as shown in in Fig.\ref{fig2:ion-wake-beam} .} This figure shows that the bubble which is excited at the earliest time undergoes phase-mixing over as several cycles. The intermediate stages of the extent of phase-mixing can be inferred from the bubbles that are closer to the beam. Here the beam is located between 170 and 180$\frac{c}{\omega_{pe}}$ (easily inferred from the radial electric field). (a) electron density in 2D cylindrical real-space, (b) corresponding longitudinal electric field profile and (c) corresponding radial-field profile. The beam-plasma parameters are the same as in Fig.\ref{fig2:ion-wake-beam} but the electron-wake is shown at an earlier time $t = 150\omega_{pe}$.}
\label{fig3:bubble-train}
\end{figure}

In this first-order approximation the right-hand side of the eq.\ref{ion-wave-equation-first-order} shows two separate timescales of the ion-wake. At earlier times, the first term on the right-hand side dominates. This is the {\it formation} phase of the ion-wake where the bubble electron oscillations undergo ordered radial motion and the bubble radial electric field excites the inertial response of the ions. This leads to the formation of the on-axis and $R_B$ ion density spikes. At later times after the phase-mixing between radial oscillators the electrons thermalize and ${\bf E}_{wk}({\bf r},t)\sim 0$. This is the {\it propagation} phase where the electron thermal pressure gradient drives the cylindrical soliton at $R_B$ radially outwards. 

The PIC simulations show the two phases of the electron dynamics. The coherent and phase-mixing phases of the electron oscillations are evident in Fig.\ref{fig3:bubble-train}(a). The electron motion in the first oscillation bucket is coherent whereas several oscillations behind the driver, the electron begin to de-cohere. This is also evident in the time-evolution of the ion soliton in Fig.\ref{fig6:ion-motion-lineouts}(b) where the radial electric field goes to zero around 200$\omega_{pe}^{-1}$ and the ion-soliton is seen moving outwards radially in \ref{fig6:ion-motion-lineouts}(a).

\subsection{Second-order non-linear ion-acoustic wake}

The linearized ion-acoustic wave equation is however inadequate to describe the propagating solitary density spike at the ion-wake edge, with ion density accumulation many times the background density. In the linear regime the homogenous ion-acoustic wave equation predicts sinusoidal radial ion oscillations that support the wave. 

When the density in the ion perturbation begins to rise to the order of the background density, the electrostatic potential due to charge-separation between the ions and the thermal electrons correspondingly rises. This leads to {\it wave-steepening} due to the preferential acceleration of ions in the direction of the ion-acoustic wave velocity. When the potential of the wave is large enough the ions get trapped and co-propagate with the ion-acoustic wave phase velocity, this non-linearity is the basis of the soliton. It should be noted that the linearized kinetic theory does not formally incorporate the trapping of particles at the wave phase-velocity. In this limit the density perturbation shape is therefore not sinusoidal as the co-propagating background ions accumulate and their density perturbation takes the form of an ion-soliton. The co-propagating ion velocity in the soliton can therefore exceed the ion-acoustic phase velocity, $v_i > c_s$ and $\mathcal{M}-1 > 0$ where $\mathcal{M} = v_i/c_s$ is the Mach number. Therefore, non-linear acoustic waves are in the form of a soliton and propagate faster than the ion-acoustic velocity. 

In the second-order, the non-linear ion-density spike $n_{i}(r,t) > n_0$ propagation is governed by the Korteweg-de Vries (KdV) equation \cite{KdV-non-linear-ion-waves} which has propagating solutions of the form $\mathcal{U}(r-\mathcal{M}c_st)$ \cite{Berezin-Karpman-1964} where $\mathcal{U}$ is the ion-acoustic waveform, a soliton solution and $\mathcal{M}$ is the Mach number of the propagating solution. Higher-order contributions to the KdV equation have also been considered. However, the standard form of the KdV equation assumes an isothermal plasma whereas the bubble-wake phase-mixes into a plasma with radial temperature gradient. In a non-isothermal plasma the effect of trapped electrons in the ion-soliton have been considered using the Bernstein-Greene-Kruskal model at the ion-acoustic velocity \cite{Schamel-trapped-particle-modes}.

It is also known that a single ion-soliton under the appropriate conditions can break-up into multiple solitons leading to a N-soliton solution. This is also a phenomenon we observe in the simulations shown in the ion density of the beam-driven case at $z = 60\frac{c}{ \omega_{pe} }$ in Fig.\ref{fig2:ion-wake-beam}.

\begin{figure}[ht!]
	\begin{center}
   	\includegraphics[width=\columnwidth]{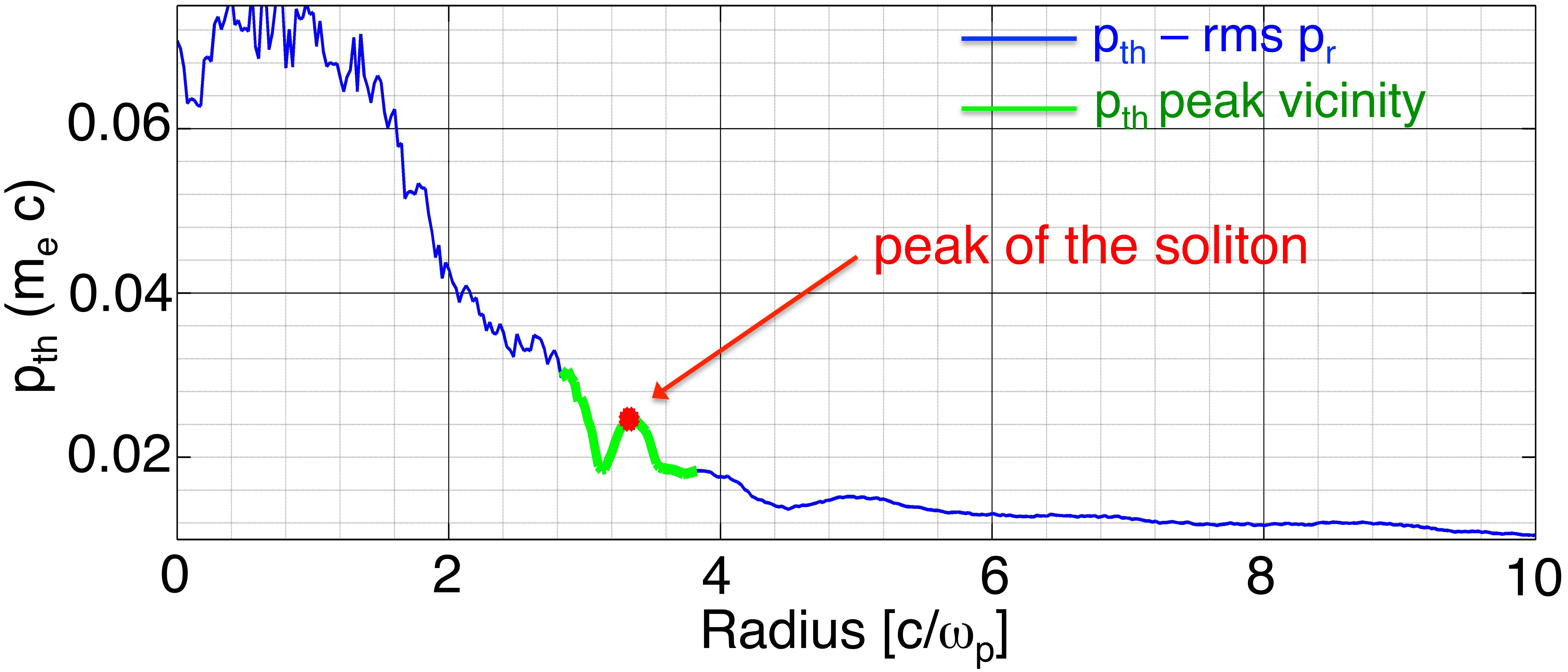}
	\end{center}
\caption{ \scriptsize {\bf Radial profile of the root-mean-square radial electron momentum (proportional to the square-root of the electron temperature, $\sqrt{T_e}$) at 460 $\omega_{pe}^{-1}$ for the beam-driven ion-wake in Fig.\ref{fig2:ion-wake-beam}}. The blue curve shows the root-mean-squared radial electron momentum, $p^e_{th}(r) = \sqrt{ \left[ \Sigma_k ~ p^2_r(k,r) n_e(k,r) \right] / \Sigma_{k} n_e(k,r) }$, profile of the wake electrons corresponding to the time in Fig.\ref{fig2:ion-wake-beam} at 460 $\omega_{pe}^{-1}$. This represents the square-root of the electron temperature, $p^e_{th} ~ \propto ~ \sqrt{T_e}$. The radial gradient of the temperature, $\frac{ \partial }{ \partial r } T_e$ is thus computed at the peak of the soliton (red) and in its vicinity (green). It is interesting to note that $\frac{\partial }{\partial r}T_e^{(1)}\biggr\rvert_{peak} = 0$. }
\label{fig4:elec-temperature-profile}
\end{figure}

We consider a description of the non-linear cylindrical ion-acoustic waves with a radial temperature gradient by retaining the second-order of perturbation in ion-density and fluid velocity. We also assume that the background electron trapping does not significantly modify the distribution function. We assume that the temperature does not change by several orders of magnitude in vicinity of the ion soliton. This assumption is also validated by the PIC simulations in Fig.\ref{fig4:elec-temperature-profile}.

To obtain the KdV equation in cylindrical coordinates with radial temperature gradient we normalize with respect to the local electron temperature, the radius: $\hat{r} = \frac{r}{\lambda_D}$, time: $\hat{t} = \omega_{pi}t = t\sqrt{\frac{4\pi e^2 n_0}{m_i}}$, electric field: $\hat{E} = \frac{e \lambda_{De} }{k_B T_e} E$, potential: $\phi = \frac{e}{k_B T_e} \Phi$, ion-density perturbation: $\hat{n}_i = n_i/n_0$, electron-density perturbation: $\hat{n}_e = n_e/n_0$, ion-fluid velocity: $\hat{v}=\frac{v_i}{c_s}$. Under this normalization the cylindrical coordinate equations transform as: electron Boltzmann distribution equation $\frac{\partial \hat{n}_e}{\partial \hat{r} } = -\hat{n_e} \hat{E} - \hat{n}_e \phi \frac{ \partial }{ \partial \hat{r} } ~ \mathrm{ln} T_e$, ion-fluid continuity equation $\frac{\partial }{\partial \hat{t} } \hat{n}_i + 2 \frac{ \hat{n}_i \hat{v}}{\hat{r}} + 2\frac{\partial}{\partial \hat{r}} \hat{n}_i \hat{v} = 0$, ion-fluid equation of motion $\frac{\partial }{\partial \hat{t} } \hat{v} + \hat{v} \frac{\partial}{\partial \hat{r}} \hat{v} = \hat{E}$ and the Poisson equation $\nabla^2\Phi = \frac{1}{\hat{r}} \frac{\partial}{\partial \hat{r}} (\hat{r}\hat{E})  = \hat{n}_i - \hat{n}_e$. The electric field $\hat{E}$ is both due to the thermal pressure and the radial fields of the wake, $\hat{E}_{wk} + \hat{E}_{th}$. But, in the following analysis the propagation of a non-linear ion-acoustic wave is considered, so we assume that the electron oscillations are thermalized and thus the effect of the fields of the wake is negligible, $\hat{E}_{wk} \rightarrow 0$.

We look for a propagating disturbance of $\hat{n}_e$, $\hat{n}_i$, $\hat{v}$ and $\hat{E}$ in a stationary background plasma with uniform background density $n_0$. We consider weakly non-linear ion-acoustic wave and expand all the wave quantities in the powers of $\delta = \mathcal{M} - 1$. We expand the ion-density: $\hat{n}_i = 1 + \delta n^{(1)}_{i} + \delta^2 n^{(2)}_{i} + \mathcal{O}(n^{(3)}_{i})$, electron density: $\hat{n}_e = 1 + \delta n^{(1)}_{e} + \delta^2 n^{(2)}_{e} + \mathcal{O}(n^{(3)}_{e})$, electric-field: $\hat{E} = \delta \hat{E}^{(1)} + \delta^2 \hat{E}^{(2)} + \mathcal{O}(\hat{E}^{(3)})$, electrostatic potential: $\phi = \delta \phi^{(1)} + \delta^2 \phi^{(2)} + \mathcal{O}(\phi^{(3)})$, electron temperature: $T_e = \delta T_e^{(1)} + \delta^2 T_e^{(2)} + \mathcal{O}(T_e^{(3)})$ and ion-fluid velocity: $\hat{v} = \delta \hat{v}^{(1)}_i + \delta^2 \hat{v}^{(2)}_i + \mathcal{O}(\hat{v}^{(3)}_i)$. Note that we have assumed that before the electron wake excitation the plasma is cold, $T_e^{(0)} \simeq 0$.

We transform to a moving frame of the steepened ion density perturbation using the coordinate transform $\xi=\delta^{1/2}(\hat{r} - \hat{t})$ and $\tau = \delta^{3/2}\hat{t}$. Therefore, the radial $\hat{r}$-derivative is, $\frac{\partial }{\partial \hat{r}} = \frac{\partial}{\partial \hat{\xi}} \frac{\partial \xi}{\partial \hat{r}} = \delta^{1/2} \frac{\partial }{\partial \xi}$ and the $\hat{t}$-derivative is $\frac{\partial }{ \partial \hat{t}} = \frac{\partial }{ \partial \xi}\frac{\partial \xi}{ \partial \hat{t}} + \frac{\partial }{ \partial \tau}\frac{\partial \tau}{ \partial \hat{t}} = - \delta^{1/2}\frac{\partial}{\partial \xi} + \delta^{3/2}\frac{\partial}{\partial \tau}$. Using this, $\hat{r} = \delta^{-1/2}(\xi + \delta^{-1}\tau)$ and $\frac{\partial \xi}{\partial \tau} = \frac{\partial \xi}{\partial \hat{t} } \frac{\partial \hat{t} }{\partial \tau} = -\frac{1}{\delta}$. We renormalize the electric field as, $\tilde{E} = \delta^{-1/2}\hat{E}$. Note that in the moving frame the potential gradient is, $E = -\frac{\partial }{\partial \hat{r}} \Phi = -\delta^{1/2} \frac{\partial }{\partial \xi} \Phi $, so $\tilde{E}$ is a more appropriate quantity.

Under the assumption that in the moving-frame the quantities of the disturbance change with small $\delta = \mathcal{M} - 1$, the terms in equations are perturbatively expanded and the terms with same powers of $\delta$ are collected. From the $\delta^{1}$ order terms of all the equations above, we infer $\Phi^{(1)} = n^{(1)}_{e} = v^{(1)} = n^{(1)}_{i}\equiv \mathcal{U}$ and $\frac{\partial }{\partial \xi}\mathcal{U} = -\tilde{E}^{(1)}$. 

Collecting the $\delta^{2}$ terms from the Boltzmann's equation we obtain $\tilde{E}^{(2)} = -\frac{\partial }{\partial \xi} n^{(2)}_{e} + \mathcal{U}\frac{\partial }{\partial \xi}\mathcal{U} - \mathcal{U}\frac{\partial }{\partial \xi}T_e^{(1)}$. From the ion-fluid equation of motion we obtain $\frac{\partial }{\partial \xi} \hat{v}^{(2)} - \frac{\partial }{\partial \xi} n^{(2)}_{e} = \frac{\partial }{\partial \tau} \mathcal{U} + \mathcal{U} \frac{\partial }{\partial \xi} T_e^{(1)}$ and from the Poisson equation we obtain $\frac{\partial^3 }{\partial \xi^3} \mathcal{U} = - \frac{\partial }{\partial \xi}(n^{(2)}_{i} - n^{(2)}_{e})$. Taking the $\delta$-order terms of the continuity equation and substituting $\mathcal{U}$ we obtain, $\mathcal{U} + \tau \left( \frac{\partial }{\partial \tau} \mathcal{U} + 2\mathcal{U}\frac{\partial }{\partial \xi}\mathcal{U} + \left[ \frac{\partial }{\partial \xi} v^{(2)} - \frac{\partial }{\partial \xi} n^{(2)}_{e} \right] \right) - \delta \left(U^2 + v^{(2)} \right) = 0$.  Neglecting quantities with $\delta$ times the second-order terms and using the $\frac{\partial }{\partial \tau} \mathcal{U}$ result above, $\frac{\mathcal{U}}{\tau} + 2\frac{\partial }{\partial \tau} \mathcal{U} + 2\mathcal{U}\frac{\partial }{\partial \xi}\mathcal{U} + \left[\frac{\partial }{\partial \xi} n^{(2)}_{e} - \frac{\partial }{\partial \xi} n^{(2)}_{i}\right] = - \mathcal{U}\frac{\partial }{\partial \xi}T_e^{(1)}$. Using the $\delta^2$ terms of the Poisson equation we obtain the driven Korteweg-de Vries equation in cylindrical coordinates \cite{Maxon-cyl-soliton}, 

\begin{align}
%
\nonumber & \Phi^{(1)} = n^{(1)}_{e} = v^{(1)} = n^{(1)}_{i}\equiv \mathcal{U} \\
& \frac{\mathcal{U}}{\tau} + 2\frac{\partial }{\partial \tau} \mathcal{U} + 2\mathcal{U}\frac{\partial }{\partial \xi}\mathcal{U} + \frac{\partial^3 }{\partial \xi^3} \mathcal{U} = - \mathcal{U}\frac{\partial }{\partial \xi}T_e^{(1)}
\label{Cylindrical-KdV-equation}
\end{align}

It differs from the cartesian-KdV equation by the term $\frac{\mathcal{U}}{\tau}$ and the driver term $- \mathcal{U}\frac{\partial }{\partial \xi}T_e^{(1)}$. The cartesian KdV equation can be analytically solved to obtain two classes of solutions: (a) self-similar solutions which are shown in \cite{Berezin-Karpman-1964} to be Airy functions and (b) soliton solutions. A \enquote{soliton} is a single isolated pulse which retains its shape as it propagates at some velocity, $v_{soliton}$. This means that for a soliton-like solution the $\mathcal{U}$ only depends upon the soliton-frame variable, $\zeta = \xi - \mathcal{M}c_s \tau$ and not on space-like $\xi$ and time-like $\tau$ variables separately. The solution of the cartesian KdV equation in this co-moving frame is $\mathcal{U}(\zeta)=3 v_s ~ \mathrm{sech}^2(\sqrt{ \frac{v_s}{2} } \zeta)$ \cite{Berezin-Karpman-1964}.

The cKdV equation and the driven cKdV equation obtained here cannot be solved analytically. However, the numerical analysis and experimental verification \cite{cyl-soliton-observation} of the cylindrical-KdV (cKdV) equation show that it supports functions of the form $\mathcal{U} \propto \mathrm{sech}^2(r-\mathcal{M}c_st)$ in the form of a cylindrical ion-soliton. But, the amplitude of the cylindrical soliton changes as it propagates. The velocity of the soliton in the cylindrical case is higher than in cartesian case \cite{Maxon-cyl-soliton}. Since the ion-wake is excited in a non isothermal plasma its velocity changes as it is driven. The mean electron temperature reduces because the volume of the ion-wake expands as the soliton moves radially outwards. It is also known to support N-soliton solution. We computationally seek the dependence of the non-linear ion-density spike on $(r-\mathcal{M}c_st)$-coordinate. 

It should be noted that such soliton solutions are supported under certain limiting condition on the Mach number, $\mathcal{M}$. The strict condition on the existence and stability of an ion-soliton arises from the requirements on the magnitude of the potential required for trapping the background ions for a given velocity of the soliton. 

Here we find that the speed of ion soliton is approximately equal to the ion-acoustic speed using the mean temperature. As this is not an isothermal plasma there is no well-defined ion-acoustic speed. So, the ion-acoustic wave is phase-mixed as its phase-velocity is spatially varying. However, we confirm from the simulations that this condition is strongly satisfied as $\mathcal{M} \geq 1$. In this scenario the mean electron temperature provides an estimate of the sound-speed. The local electron temperature of the ion soliton as shown in Fig.\ref{fig4:elec-temperature-profile}, is used to calculate Mach number, $\mathcal{M}$ and thus a stability criteria can be derived. This problem is represented using the condition on the Sagdeev psuedo-potential, $\mathcal{V}(\phi) = - \left( \mathrm{exp}(\phi) -1 + \mathcal{M}(\mathcal{M}^2 - 2\phi)^{1/2} - \mathcal{M}^2 \right)$ that it has to be a real number. This condition is satisfied when $\mathcal{M}^2 - 2\phi \geq 0$ therefore, $\phi < \mathcal{M}^2/2$ and $\phi_{max} = \mathcal{M}^2/2$. Using this we find the well-known condition, 

\begin{align}
%
\nonumber & 1 < \mathcal{M} < 1.6, \quad v_i < 1.6 ~ c_s \\
& \Phi < \frac{\mathcal{M}_{max}^2}{2} = 1.28 
\label{KdV-solution-existence}
\end{align}

As will be shown later, we find from simulations that the Mach number calculated using the mean temperature is well within these bounds and thus the soliton is stable.


\section{Non-linear Ion-wake model: \\ ion-soliton excitation phase}
\label{ion-soliton-excitation-phase}

\begin{figure}[ht!]
	\begin{center}
   	\includegraphics[width=\columnwidth]{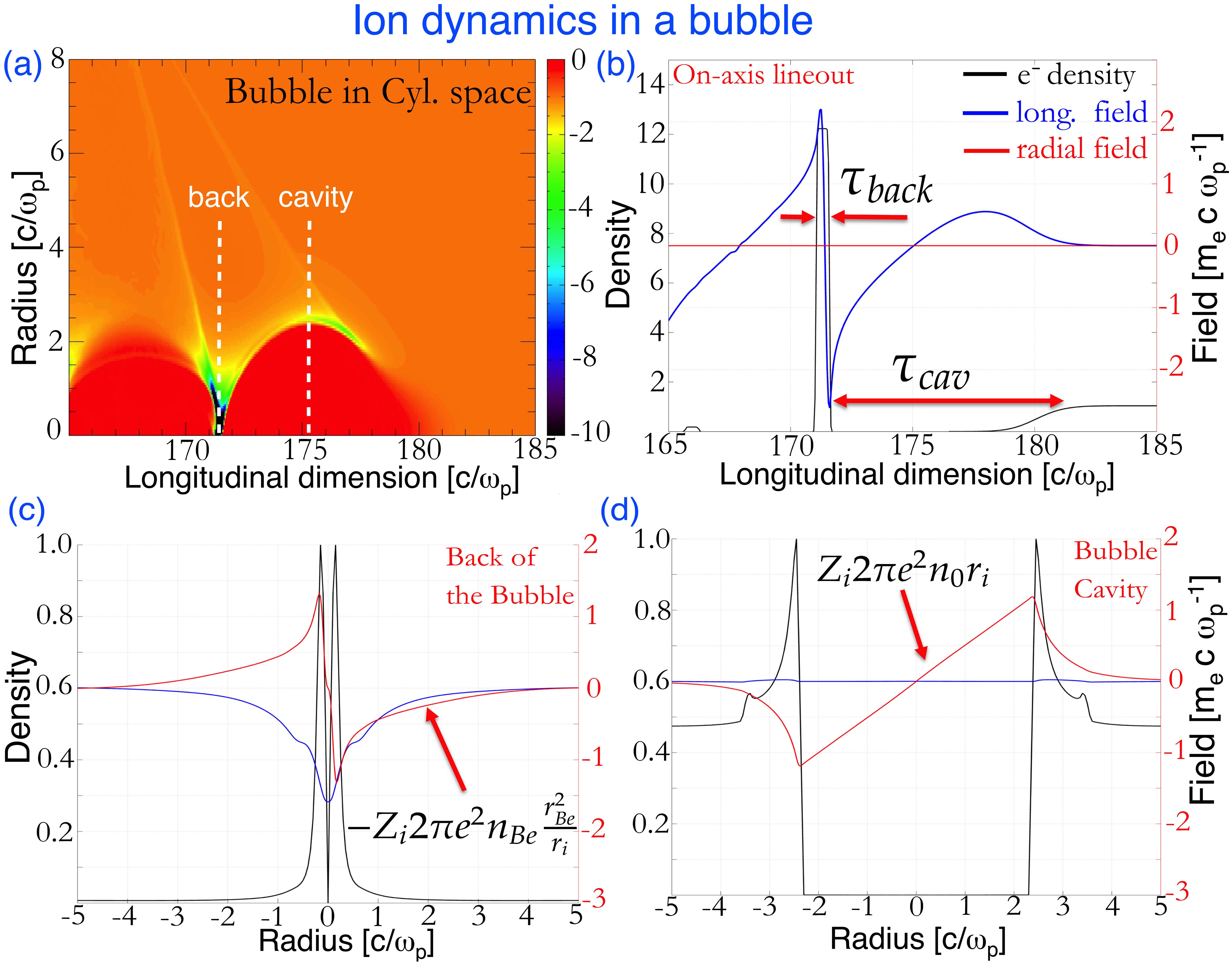}
	\end{center}
\caption{ \scriptsize {\bf Ion dynamics in longitudinally asymmetric phases of the radial forces in an electron bubble}. (a) electron density of a bubble in 2D cylindrical real-space. (b) longitudinal on-axis profile of the electron density (black), longitudinal field (blue), focusing field (red). (c) radial-field profile close to the back of the bubble. This is the focussing ``suck-in" phase for the ions. (d) the fields at the center of the ion-cavity of the bubble. This is the defocussing ``push-out" phase for the ions. } 
\label{fig5:bubble-ion-dynamics}
\end{figure}

Since the characteristic time of ion-motion is much longer than the electron oscillations, the longitudinal field ${\mathbf E}_{wk}\cdot\hat{z}$ averages out over the full bubble electron oscillation. So, the ions do not gain any net longitudinal momentum. However, atypical radial ion-dynamics arise because the radial fields, ${\mathbf E}_{wk}\cdot\hat{r}$ are asymmetric in time as shown in Fig.\ref{fig5:bubble-ion-dynamics} and do not average to zero, driving an average radial ion-momentum. 

The first stage of the ion-wake formation is controlled by the different time-asymmetric phases in Fig.\ref{fig5:bubble-ion-dynamics} of the bubble radial impulses namely, ``suck-in" due to the electron compression in the back of the bubble $F^{back}$ during $\tau_{back}$ shown in Fig.\ref{fig5:bubble-ion-dynamics}(c), and the ``push-out" due to the mutual-ion space-charge Coulomb repulsion force $F^{sc}$ during $\tau_{cav}$ shown in Fig.\ref{fig5:bubble-ion-dynamics}(d). The suck-in force is spatially-periodic at non-linear plasma wavelength, $\lambda_{Np} \approx 2R_B$ with a duty-cycle $\mathcal{D}=\frac{\tau_{back}}{\tau_{back}+\tau_{cav}}\ll 1$. In addition to the plasma wake, the propagating energy sources themselves impart impulses such as the laser ponderomotive force $F^{pm}\tau_{las}$ where $F^{pm}_{e}(r,z) = -\frac{m_ec^2}{2\gamma_e}{\bf \nabla}_{r,z}|{\bf a}(r)|^2$ and the radial force of the drive beam $F_{b}\tau_{b}$ where $F_{b}(r)=- 2\pi e^2 n_{b} r$. We neglect the driver impulses (below threshold intensity for direct non-linear ion excitation \cite{ion-motion-intense-beam}\cite{ion-motion-beam-emittance})) because they act on the ions over their sub-wavelength short duration. This is unlike the slowly-propagating wake-plasmon bubbles that undergo continual interaction over many plasma periods. The validity of this assumption is evident from the laser ion-wake in Fig.\ref{fig1:ion-wake-laser}. Since the ponderomotive force of a laser driver is an outward force for both the electrons and ions, the on-axis density-spike cannot be from this force. Similarly the ion-density-spike at the radial wake-edge in an electron beam driven ion-motion cannot be excited directly by the force of the beam, and is caused by the electron wake's radial-edge density compression.

The Lagrangian fluid model of the ions in a bubble consists of ion-rings under cylindrical symmetry with $m_{i} d^2r_{i}/dt^2 = \Sigma F_{wk}$. The bared-ion region inside the bubble is assumed to be a positively charged cylinder under steady-state approximation ($R_{B} > r_{Be}$, back of the bubble electron compression radius). The force on the ions from the non-linear electron compression $\delta n_e = n_{Be}\gg n_0$ in the back of the bubble and radius $r_{Be}$, pulls the ion rings inward; and within the bubble, the space-charge force of the ions opposes it and prevents full collapse. The \enquote{suck-in} force on ions is $F^{back} = - Z_{i} 2\pi e^2 n_{Be} \frac{r^2_{Be}}{r_{i}}$. The space-charge force on the ions in the cavity is $F^{sc} = Z_{i} 2 \pi e^2 n_0 r_{i}$. The equation of motion is $m_{i}d^2r_{i}/dt^2 - \frac{c\beta_{\phi}}{\lambda_{Np}}(F^{sc}\tau_{cav} - F^{back}\tau_{back}) = 0$ using, $\omega_{pi}^2 = Z_{i} ~ 4\pi e^2 n_0 / m_{i}$, we have, 

\begin{align}
\frac{d^2r_{i}}{dt^2} + \beta_{\phi} \frac{\omega^2_{pi}}{2}\left(\frac{n_{Be}}{n_0} \frac{\tau_{back}}{\tau_{cav}} \frac{r^2_{Be}}{r_{i}^2} - 1 \right) r_{i} = 0
\label{ion-ring-motion}
\end{align}

where we have assumed that $c\tau_{cav}/\lambda_{Np} \simeq 1$. Therefore the equilibrium radius where the impulses balance out is $r_{i}^{eq}=r_{Be}\sqrt{\frac{n_{Be}}{n_0} ~ \mathcal{D} }$. The ion-rings at $r_{i} \le r_{i}^{eq}$ collapse inwards towards the axis resulting in an on-axis density spike. Whereas the ion-rings at $r_{i} \ge r_{i}^{eq}$ move out away from the axis. For $m_{i}/Z_{i}>m_{p}$ the ion-response is slower but similar.

When the radially outward moving ion-rings reach beyond $R_B$, there is excess net negative charge of the wake electrons within the bubble-sheath. As a result the radially propagating ion rings get trapped and start accumulating just inside the bubble and cannot freely move beyond, forming a density compression at $R_B$. So, the cylindrical ion soliton is formed around $R_B$. This is seen in Fig.\ref{fig1:ion-wake-laser} and Fig.\ref{fig2:ion-wake-beam} where the ion and electron density peak at $R_B$. 

In the non-linear regime $R_B\gg c/\omega_{pe}$, the spatial-scale of the ion-wake is over several $c/\omega_{pe}$. This is due to the balance between the opposite radial forces on the electrons at $R_B$, from the driver and the ion cavity \cite{Pukhov-laser-bubble}\cite{Lu-bubble-regime}. In the laser-driven bubble $F^{pm}_{las} = \frac{m_ec^2}{2\gamma_e}{\bf \nabla}_r \lvert{\bf a}_0(r)\rvert^2 \simeq F_{cav} = 2\pi e^2 n_0 R_{B}$ gives $R_{B} \sim (c/\omega_{pe})^2 \frac{1}{\gamma_e} {\bf \nabla}_r \lvert{\bf a}_0(r)\rvert^2$ when simplified using ${\bf \nabla}_r|{\bf a}_0(r)|^2 \simeq a_0^2/R_{B}$ and $\gamma_e \simeq a_0$, $R_{B} \simeq \frac{c}{\omega_{pe}} \sqrt{a_0}$ (computationally, $\simeq2\sqrt{a_0} ~ c/\omega_{pe}$ \cite{Lu-bubble-regime}). In the beam-driven bubble $F_{b}(R_B) = 2\pi e^2 n_b r_b^2 / R_B \simeq F_{cav} = 2 \pi e^2 n_0 R_B$ which gives, $R_B \simeq \sqrt{\Lambda_b/(\pi n_0)}$, where $\Lambda_b = n_b \pi r_b^2$ is the line charge density of the beam.

We verify the above model using $2\frac{1}{2}D$ OSIRIS PIC simulations \cite{osiris-code-2002} of the ion-wake in the bubble regime by simulating various laser-pulses in cartesian coordinates and electron-beams in cylindrical coordinates. We initialize Eulerian specification of the plasma (non-moving window) of pre-ionized $Z_i=1$ homogeneous density with a density-ramp over 20 $c/\omega_{pe}$ at each of the vacuum-plasma interfaces. We resolve the smallest spatial scale, $c/\omega_{pe}$ in the beam case and $c/\omega_0$ in the laser case (laser frequency $\omega_0$), with 20 cells. The time-step satisfies causality and minimizes numerical dispersion. It is normalized to $\omega_{pe}^{-1}$ for beam and to $\omega_0^{-1}$ for lasers. We use absorbing boundary conditions for fields and particles. The laser pulse is circularly polarized with radially Gaussian and longitudinally polynomial profile \cite{laser-pulse-profile} with $a_0=4$ (not shown $a_0=1.0$ to $40.0$), pulse length of $30\frac{c}{\omega_0}$, matched focal spot-size radius of $40\frac{c}{\omega_0}$, and laser frequency $\omega_0=10\omega_{pe}$. The electron beam is initialized with $\gamma_b\sim38,000$, $n_b=5n_0$ (not shown $n_b=0.25n_0$ to $50n_0$) and spatial Gaussian-distribution with $\sigma_r=0.5\frac{c}{\omega_{pe}}$ and $\sigma_z=1.5\frac{c}{\omega_{pe}}$.

We compare the electron-beam driven ion-wake soliton structure in theory to the simulations in Fig.\ref{fig5:bubble-ion-dynamics}(a) and \ref{fig3:bubble-train}(a). The observed $R_B=2.45 ~ c/\omega_{pe}$ (just behind the beam) whereas the estimated bubble radius is $R_B = \sqrt{n_b/n_0 ~ (2.3\sigma_r)^2} = 2.57 ~ c/\omega_{pe}$ ($r_b\simeq 2.3\sigma_r$, the assumption $r_b\ll R_B$ is not well satisfied). In Fig.\ref{fig2:ion-wake-beam} which is in the propagation-phase, the observed ion-wake soliton is located at $r \simeq 3.3 ~ c/\omega_{pe}$ at 460 $\omega_{pe}^{-1}$ which is about $1.7 \frac{2\pi}{\omega_{pi}}$. The ion-soliton is excited at an early time around $R_B$ and in the snapshot in Fig.\ref{fig2:ion-wake-beam} it has propagated outwards. The on-axis density spike drops to a minimum at $r_{i}^{eq} \approx 0.45c/\omega_{pe}$ in Fig.\ref{fig2:ion-wake-beam} whereas the estimated $r_{i}^{eq} = 0.5 ~ c/\omega_{pe}$ ($n_{Be}/n_0 \simeq 12$, $\mathcal{D} \simeq 0.1$, $r_{Be}\simeq0.5 ~ c/\omega_{pe}$). The radial ion momentum $p_r-r$ phase-space in Fig.\ref{fig7:radial-phase-spaces}(b) shows the ions accumulate at the axis and the channel edge at a time corresponding to Fig.\ref{fig2:ion-wake-beam}(b). The ions at the channel edge are seen to have a drift velocity and a thermal spread. The radial electron momentum $p_r-r$ phase-space in Fig.\ref{fig7:radial-phase-spaces}(a) shows that a large density of thermalized electrons are trapped within the ion soliton which is confirmed from the density plots in Fig.\ref{fig2:ion-wake-beam}(a).

In the laser-driven bubble simulations the expected and observed $R_B \simeq 4 ~ c/\omega_{pe}$ as shown in Fig.\ref{fig1:ion-wake-laser}(a). In Fig.\ref{fig1:ion-wake-laser}(c) the ion-wake soliton is created at  $r=4.2 ~ c/\omega_{pe}$. The expected and observed on-axis density-spike radius is $r_{i}^{eq} = 0.45 ~ c/\omega_{pe}$ ($n_{Be}/n_0 \simeq 8$, $\mathcal{D} \simeq 0.1$, $r_{Be}\simeq0.5 ~ c/\omega_{pe}$). The model for the excitation of this structure of the non-linear wake has been verified for a range of laser and beam parameters from quasi-linear to strongly non-linear electron wake regime.

\section{Non-linear Ion-wake model: \\ ion-soliton propagation phase}
\label{ion-soliton-propagation-phase}

As described in section \ref{ion-soliton-excitation-phase} the electron bubble-wake train fields excite a cylindrical ion soliton. Eventually, the electron oscillations phase-mix and thermalize as electron thermal energy on the time-scale of about an ion plasma period. The cylindrical soliton then propagates radially outwards driven by the temperature gradient as shown in eq.\ref{Cylindrical-KdV-equation}. This is the modified cKdV equation in a non-equilibrium condition when an electron temperature gradient also contributes to the force on the ions.

\begin{figure}[ht!]
	\begin{center}
   	\includegraphics[width=4.8in]{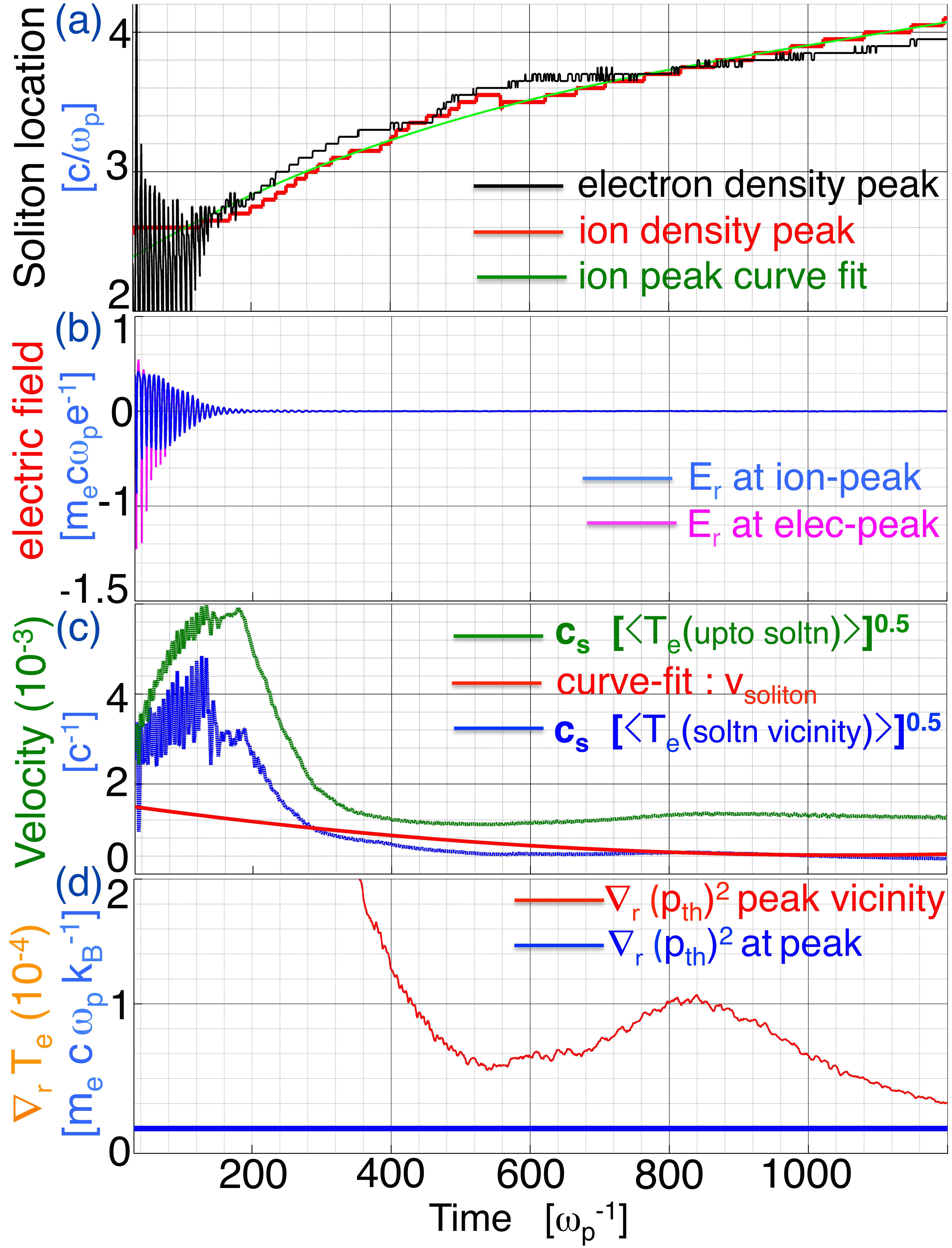}
	\end{center}
\caption{ \scriptsize {\bf Time evolution of the cylindrical ion soliton}. (a) electron (black) and ion (red) spike radial positions (in terms of $c\omega_{pe}^{-1}$) with time and a third-order fit (green) for the position of the ion density-spike of the soliton. (b) radial wakefields of the electron bubble oscillations (in terms of $m_e c \omega_{pe}e^{-1}$) at the electron density spike (magenta) and at the ion density spike (blue). (c) radial velocity of the ion density spike of the soliton calculated from the third-order fit curve (red). An estimate of the sound speed (green) using the mean temperature, between the axis \& the soliton location (green) and in the vicinity of the soliton peak (blue), in the expression $c_s = \sqrt{k_B \langle T_e \rangle / m_i}$. Since the plasma is not isothermal the mean temperature is calculated by averaging the temperature of electrons over the indicated spatial region. (d) gradient of the electron temperature at the soliton ion density peak (blue) and in the vicinity of the peak (red). The vicinity of the ion density peak of the soliton is defined as shown in Fig.\ref{fig4:elec-temperature-profile}. }
\label{fig6:ion-motion-lineouts}
\end{figure} 

The channel-edge density spike, with a form similar to the cKdV-solution in the $r-\mathcal{M}c_st$ frame as shown in Fig.\ref{fig1:ion-wake-laser} and Fig.\ref{fig2:ion-wake-beam} to be propagating radially outwards. The propagation phase starts around $t = 200 \omega_{pe}^{-1}$ as the radial electric fields $\mathbf{E}_{wk} \rightarrow 0$ as shown in Fig.\ref{fig6:ion-motion-lineouts}(b). The propagation phase is evident in Fig.\ref{fig6:ion-motion-lineouts}(a) where the red curve is the position of the peak of the ion-soliton in time. The cylindrical ion-soliton has propagated from  $r_{soliton}(460 \omega_{pe}^{-1}) = 3.3 c/\omega_{pe}$  (also seen in Fig.\ref{fig2:ion-wake-beam}) to $r_{soliton}(1100 \omega_{pe}^{-1}) = 4.1 c/\omega_{pe}$ which corresponds to an average speed of $\langle v_{soliton} \rangle = 0.0013c$. 

We compare the time-averaged soliton speed $\langle v_{soliton} \rangle$ to the average speed of sound, $c_s/c = p^{e}_{th} \sqrt{\frac{\Upsilon}{2}\frac{m_e}{m_i}}$ where the average $p^{e}_{th} \simeq 0.06$ from the electron phase-space (not shown). This gives $c_s \simeq 0.001c$ ($\Upsilon = 2$ for 2D) in agreement with the average soliton velocity. Using this time-averaged analysis we see that $\mathcal{M} \simeq 1.3$ and so the stability criteria in eq.\ref{KdV-solution-existence} is satisfied. 

However, as the soliton moves out the volume between the axis and the soliton edge increases thus the electron thermal energy spreads over a larger volume. The re-distribution of the thermal energy over a larger volume leads to the reduction in the temperature with time. The soliton is not freely propagating but is driven by the radial gradient of the electron temperature as shown in eq.\ref{Cylindrical-KdV-equation}. The soliton speed thus changes in time as shown in the red curve of Fig.\ref{fig6:ion-motion-lineouts}(c). The sound speed also varies with time and it is estimated using the temperature at that instant using, $c_s(t) = \sqrt{k_B \langle T_e(t) \rangle / m_i}$ in terms of the root mean squared electron radial momentum. As the plasma is not in thermal equilibrium and its temperature varies radially as shown in Fig.\ref{fig4:elec-temperature-profile}. The root-mean-square radial momentum represents the temperature at an instant of time and is calculated over radial dimension - from the axis to the soliton location: channel-$\langle p^e_{th} \rangle = \sqrt{ \left[ \Sigma^{r_{sol}}_{r=0} ~ p^2_r(r) n_e(r) \right] / \Sigma^{r_{sol}}_{r=0} n_e(r) } ~ \propto ~ \sqrt{T_e}$ or in the vicinity of the soliton: soliton-$\langle p^e_{th} \rangle = \sqrt{ \left[ \Sigma^{r_{sol}+\epsilon}_{r_{sol}-\epsilon} ~ p^2_r(r) n_e(r) \right] / \Sigma^{r_{sol}+\epsilon}_{r_{sol}-\epsilon} n_e(r) } ~ \propto ~ \sqrt{T_e}$ from the $p_r-r$ phase-space. The instantaneous sound speed, $c_s(t)$ computed with channel-$\langle p^e_{th} \rangle$ is in the green curve in Fig.\ref{fig6:ion-motion-lineouts}(c) and $c_s(t)$ computed with soliton-$\langle p^e_{th} \rangle$ is in the blue curve in Fig.\ref{fig6:ion-motion-lineouts}(c). The extent of $\epsilon$ around the soliton peak is shown in Fig.\ref{fig4:elec-temperature-profile}.

We compare the curves in (i) red: $v_{soliton}(t)$ (from the 3rd-order curve-fitting to the position of the ion-density peak), (ii) green: $c_s(t)$ from channel-$\langle p^e_{th}(t) \rangle$ and (iii) blue: $c_s(t)$ from soliton-$\langle p^e_{th}(t) \rangle$ in Fig.\ref{fig6:ion-motion-lineouts}(c). From the comparison it is observed that they are in good agreement. It can be seen that the velocity of the soliton estimated using the location of the ion-density peak (red) lies between $c_s(t)$ calculated using the average temperature over the channel (green) which is the upper limit and $c_s(t)$ calculated using the average temperature over the soliton (blue) which is the lower limit. Note, that since $v_{soliton}(t)/c_s = \mathcal{M}(t) > 1$ the blue curve which is lower than the $v_{soliton}$ curve in red, is a better estimate of the sound speed in the vicinity of the ion-soliton. The sound-speed with radial momentum averaged over the channel is over-estimated because the thermal electron spread estimation also includes the velocity of the electrons drifting with the ion-soliton.

We also present the radial gradient of the electron temperature, $\frac{ \partial }{ \partial r } T_e(r,t)$ in Fig.\ref{fig6:ion-motion-lineouts}(d). It is interesting to note from the blue curve in Fig.\ref{fig6:ion-motion-lineouts}(d) that the temperature gradient at the peak of the ion-soliton is zero, $\frac{ \partial }{ \partial r } T_e(r,t) \big\rvert_{peak} = 0$. In the vicinity of the soliton peak we see that the gradient of the temperature follows the variation in the ion soliton velocity, this follows from eq.\ref{Cylindrical-KdV-equation}. The vicinity of the soliton peak is shown as the green curve overlaid on the thermal momentum curve in Fig.\ref{fig4:elec-temperature-profile}.

In Fig.\ref{fig2:ion-wake-beam}(b) N-soliton formation is observed in the ion-density at around $z \simeq 60 \frac{c}{\omega_{pe}}$. The single-ion soliton is seen splitting into several solitons. The N-soliton solution is formed because different ion-rings that are driven in the ``push-out" phase have a radial position dependent defocussing force acting on them, $F^{sc}(r_{i}) = Z_{i} 2 \pi e^2 n_0 r_{i}$. This is shown in Fig.\ref{fig5:bubble-ion-dynamics}(d). Thus the ion-rings originating at a larger radii from the axis are pushed outwards with a force of a higher magnitude and the rings originating at a smaller radii just outside $r_i^{eq}$ are pushed outwards by a smaller force. So, over a longer time the set of ion-rings with a higher initial momentum propagate radially outwards at a larger radial velocity. This break-up of a single ion-soliton into N-solitons occurs over a longer time-scale because the difference in momentum is small compared to the average momentum. 

The thermal momentum, $p_{e}^{th}$ at this time is less than one-tenth of the peak wake quiver momentum. There are several reasons for the cooling, such as, transfer of the wake energy to the ions and the trapped electrons \cite{phase-mixing-trapping}, escape of the highest energy electrons and un-trapped ions from the channel edge, energy loss to the bow-shock and the re-distribution of the energy over an expanding volume. The peak radial ion-momentum is $\simeq0.005$ which shows that not all the radially propagating ions are trapped. The un-trapped free-streaming ions at $\simeq 7c/\omega_{pe}$ can be distinguished from the ions at the channel-edge in $p_r-r$ phase-space.

\begin{figure}[ht!]
	\begin{center}
   	\includegraphics[width=4.8in]{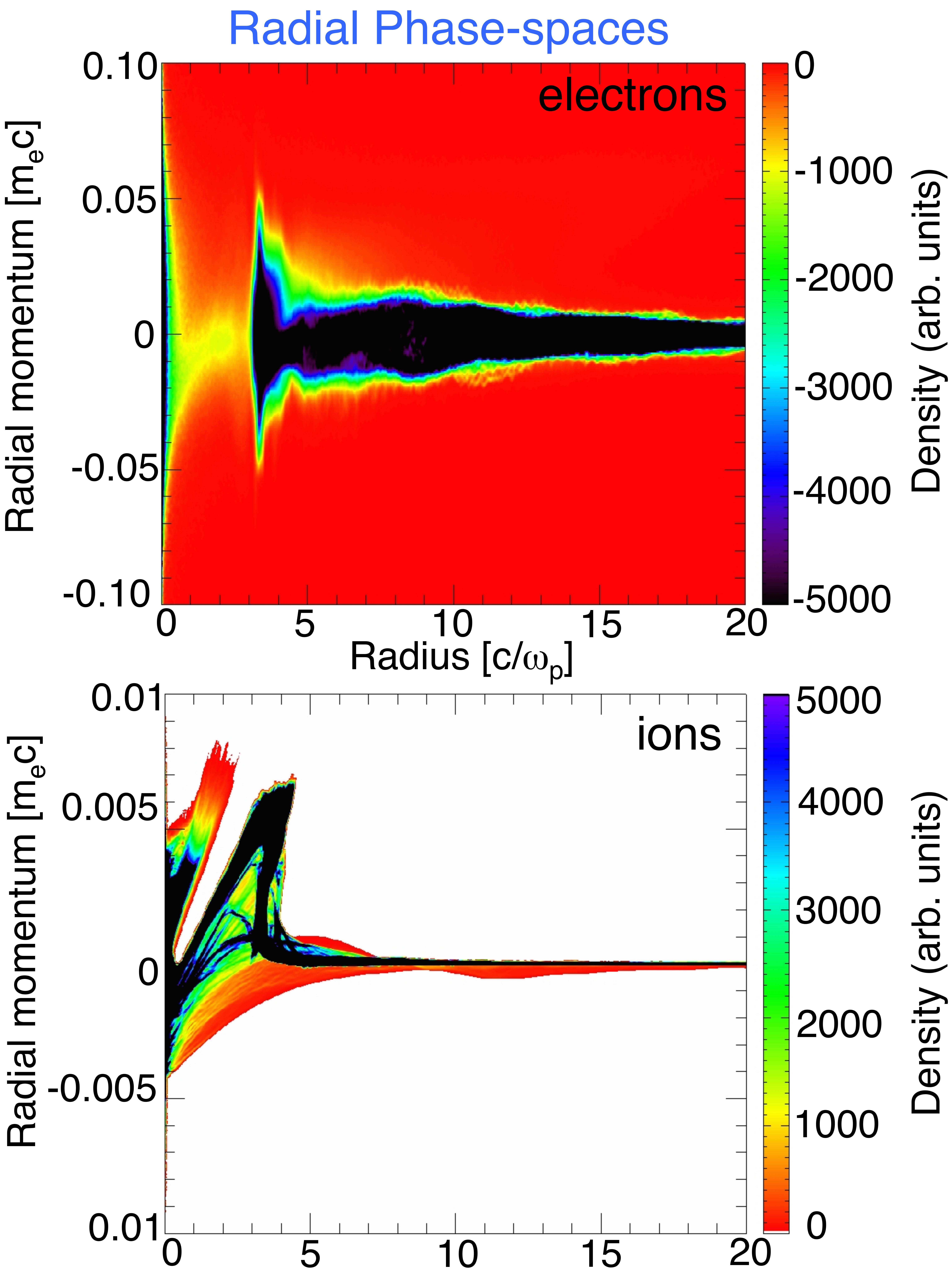}
	\end{center}
\caption{ \scriptsize {\bf Radial phase-space snapshots of the electron and ion density in Fig.\ref{fig2:ion-wake-beam}}. (a) electron $p_r-r$ radial momentum phase-space showing the accumulation of thermalized electrons within the ion-soliton. (b) ion $p_r-r$ radial momentum phase-space showing the on-axis and ion-wake edge ion accumulations. }
\label{fig7:radial-phase-spaces}
\end{figure}

\section{Positron acceleration in the \\ Ion-wake channel}
\label{ion-soliton-positron-wake}

Positron acceleration using the ion-wake channel is explored in the non-linear suck-in regime of positron-beam radii $r_{pb} > c/\omega_{pe}$ and peak density $n_{pb} > n_0$. In this regime $r_{pb} \gg r_{i}^{eq}$, so the on-axis ion density has a radially limited defocussing force. 

It is well known \cite{positron-accln-2001} that in a homogeneous plasma positron beam driven wakes have two major problems \cite{positron-IPAC-2015} - (i) The plasma electrons collapsing to the axis under the force of a positron beam start at different radii. The positron beam radial force driving the ``suck-in" decreases with the radii. As a result, the electron compression is not optimal. (ii) The plasma ions located in the path of the positron beam result in a de-focussing force on it. The transport of the positron in a positron-beam driven wake in homogenous plasma is not ideal.

As the ion-wake channel is a practical realization of the hypothesized ideal hollow-channel plasma \cite{positron-accln-2001} we examine its excitation by a positron-beam \cite{positron-IPAC-2015}. In this section we analyze whether the positron-beam driven wake-fields excited in the ion-wake can be used for the acceleration and transport of a positron beam.

We describe the analytical model of the radial electron ``suck-in" based excitation of a positron beam wake in the plasma. The equation of motion of the plasma electron rings at $r$ from the axis, under the positron beam suck-in force is 

\begin{align}
\frac{\operatorname d^2}{\operatorname d\xi^2}r \propto - \frac{1}{r} n_{bp}(\xi) r_{bp}^2(\xi)
\label{electron-ring-collapse}
\end{align}

where $\xi = c\beta_{pb}t - z$ is the space just behind the positron beam with velocity $c\beta_{pb}$ driving the collapse. This is a non-linear second-order differential equation of the form, $$r'' = f(r,r',\xi)$$ where $f$ is not linear in $r$. Under the assumption about the positron-beam properties, $n_{bp}(\xi)$ and $r_{bp}(\xi)$ being constant during the entire interaction of the positron-beam with the hollow-channel over its full length. So, upon dropping the dependence on $\xi$ the equation simplifies to its {\it special case} which has analytical solutions, $r'' = f(r,r')$. The solution to this equation is \cite{positron-IPAC-2015},

\begin{align}
r_{ch}\sqrt{\pi} ~ \mathrm{erf} \left( \sqrt{\mathrm{ln}( r_{ch} /r)} \right) = -\sqrt{2\mathcal{C}} ~ \xi
\label{positron-beam-electron-collapse}
\end{align}
 
where $\mathcal{C}=\frac{1}{2\pi\beta_b^2} \frac{n_{bp}}{n_0} \pi \left(\frac{r_{bp}}{c/\omega_{pe}}\right)^2$. Therefore the collapse time-duration is $\xi_{coll} =  -r_{ch}\sqrt{\frac{\pi}{2\mathcal{C}}} $. We note that there is an anomaly that exists in our problem formulation and the solution because we have not taken into account the space-charge force of the compressing electrons as they collapse to the axis and this force balances the suck-in force of the positron beam. Under these approximations the collapse time is 

\begin{align}
\tau_{c}=\sqrt{\pi} \frac{ r_{ch}}{\omega_{pe} \sqrt{ n_{bp}/n_0 } r_{pb} }
\label{positron-beam-electron-collapse}
\end{align} 

Also, note that we have neglected the initial expansion velocity of the channel, $dr_{ch}/dt$)\cite{positron-accln-2001}. For optimal compression avoiding phase-mixing, the electron rings should collapse over, $\tau_c \simeq \mathcal{D}\lambda_{Np}/c$ where $\lambda_{Np}$ is the non-linear wavelength of the positron-driven wake and $\mathcal{D}$ is the duty-cycle of compression phase. So, the optimal channel radius is $r_{ch}^{opt} \simeq  2\sqrt{\pi}\mathcal{D}\frac{\lambda_{Np}}{\lambda_{pe}}\frac{\omega_{pb}}{\omega_{pe}} r_{pb}$. The scaling of the $r_{ch}^{opt}$ with positron beam parameters is show in \cite{positron-IPAC-2015}.

\begin{figure}
	\begin{center}
   	\includegraphics[width=4.8in]{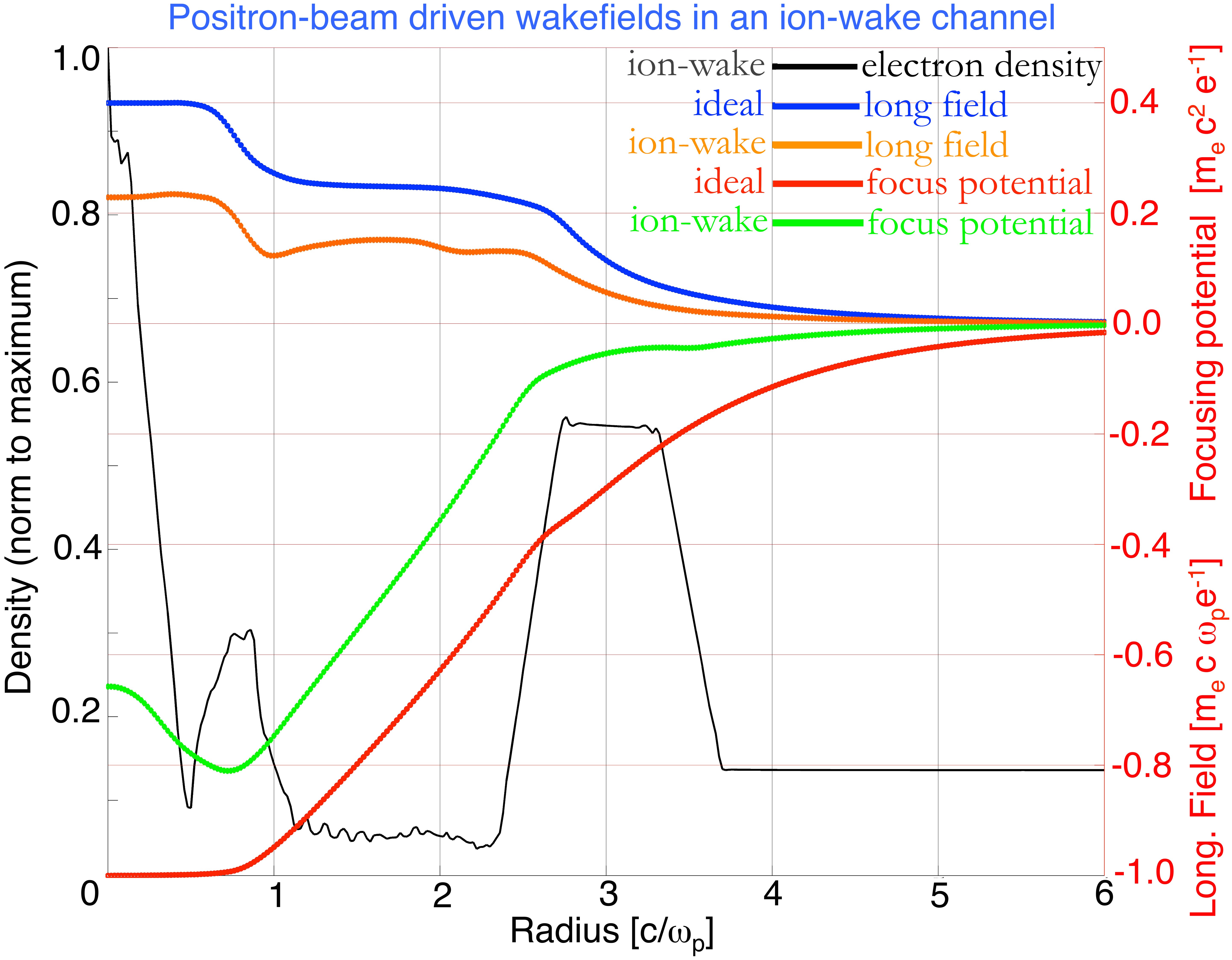}
	\end{center}
\caption{ \scriptsize {\bf Radial profile of wakefields excited by a positron-beam in ideal-channel versus an ion-wake channel}. The radial profile of the normalized electron density (black) in an ion-wake channel (normalized to the maximum electron compression) at longitudinal location of the peak accelerating wakefield ($r_{pb} = 2.3c/\omega_{pe}$, $\gamma_{pb}=38000$, $n_{pb}=1.3 n_0$). The radial profile of the accelerating-wakefield and normalized focussing-wakefield potential (radial field integrated from the edge of the box to a radius).}
\label{fig8:positron-acc-field-radial-profiles}
\end{figure}

Using 2-$\frac{1}{2}$D PIC simulations in a moving window we study the positron beam driven wakefields in cylindrical geometry. We compare positron acceleration in an ideal (Heaviside density function, $n_0\mathsf{H}(r-r_{ch})$) and an ion-wake channels (on-axis and channel-edge density-spike, channel minimum density of $0.1n_0$) with $r_{ch}= 2.5 ~ c/\omega_{pe}$. For non-linear wake parameters $r_{pb}=2.3 c/\omega_{pe}$, $n_{pb} = 1.3 n_0$ and $r_{ch}^{opt}\simeq 2.3 c/\omega_{pe}$ ($\mathcal{D}\frac{\lambda_{Np}}{\lambda_{pe}}=0.25$). Fig.\ref{fig8:positron-acc-field-radial-profiles} shows that the peak on-axis accelerating field is $0.4~m_ec\omega_{pe}e^{-1}$ for an ideal channel and $0.2~m_ec\omega_{pe}e^{-1}$ for the ion-wake channel. Fig.\ref{fig8:positron-acc-field-radial-profiles} also shows that the focussing potential (normalized to 27.6 $m_ec^2e^{-1}$) is similar and overall focussing in both cases. However, in the ion-channel the radial field is defocussing around the on-axis ion-spike. 

Thus the ion-wake channel is useful for accelerating and transporting positrons despite the lower accelerating fields in comparison with the ideal channel. We also note that the on-axis density spike has detrimental effect on the focussing fields near the axis. Ideal channels of a few $c/\omega_{pe}$ are technologically challenging whereas the ion-wake channel of radius $r_{ch}\gtrsim R_B$ is formed behind every bubble-wake.

\section{Conclusion}
In conclusion, using theory and PIC simulations we have shown the dynamics of the formation and evolution of a non-linear ion-wake excited by the well-characterized electron bubble-wakefields \cite{cavitation-laser}-\cite{cavitation-beam-expt}. We have shown that the non-linear ion-wake has the characteristic cylindrical ion-soliton solution and evolves to an N-soliton solution over longer time as described by the cKdV equation. Thus over the period of persistence of the ion-soliton a second electron bunch cannot be accelerated in the plasma. This establishes an upper limit on the repetition rate of the plasma collider. We have also shown the feasibility of using the ion-wake channel for positron acceleration in an experimentally relevant parameter regime.

\begin{acknowledgments}
Work supported by the US Department of Energy under DE-SC0010012 and the National Science Foundation under NSF-PHY-0936278. We acknowledge the OSIRIS collaboration \cite{osiris-code-2002} for providing us with the code and support for PIC simulations. We thank Dr. V. Yakimenko for useful ideas and suggestions. We acknowledge encouragement for experiments from Dr. M. Hogan and the FACET group at Stanford Linear Accelerator Laboratory and Prof. M. Downer's group at University of Texas at Austin. We acknowledge the 256-node {\it Chanakya} server at Duke university supported and maintained by Mr. Mahmood Sayed.
\end{acknowledgments}


\end{document}